\documentstyle[12pt]{article} \textwidth=6.25in \oddsidemargin=0in
\begin{document} 
\newcommand{\dl}{\delta} \newcommand{\iy}{\infty} \newcommand{\La}{\Lambda}
\newcommand{\ra}{\rightarrow} \newcommand{\tl}{\widetilde}
\newcommand{\pl}{\partial} \newcommand{\noi}{\noindent} \renewcommand{\sp}{\vskip2ex}
\newcommand{\bq}{\begin{equation}} \newcommand{\eq}{\end{equation}}
\newcommand{\th}{\theta} \newcommand{\bR}{{\bf R}} \newcommand{\bZ}{{\bf Z}} 
\newcommand{\sr}{^{(r)}} \newcommand{\bC}{{\bf C}} \newcommand{\la}{\lambda} 
\newcommand{\ph}{\varphi} \newcommand{\om}{\omega} \newcommand{\inv}{^{-1}}
\newcommand{\ov}{\over} \newcommand{\Om}{\Omega} \newcommand{\si}{\sigma}
\newcommand{\hf}{{1\ov2}} \newcommand{\ba}{\left(\begin{array}{cc}} 
\newcommand{\ea}{\end{array}\right)} \newcommand{\pxy}{{\pl^2\ov \pl x\pl y}}
\newcommand{\Px}[1]{{\pl#1\ov \pl x}} \newcommand{\Py}[1]{{\pl#1\ov \pl y}}
\newcommand{\Pxy}[1]{{\pl^2#1\ov \pl x\pl y}} \newcommand{\tr}{{\rm tr}\,}
\renewcommand{\hat}{\widehat}
\begin{center}{\large \bf An Integral Operator Solution
to the Matrix Toda Equations}\end{center}\vskip4ex

\begin{center}{{\bf Harold Widom}\\\sp
{\it Department of Mathematics\\
University of California\\ Santa Cruz, CA 95064, USA\\
e-mail address: widom@math.ucsc.edu}}\end{center}\sp

\begin{abstract}
In previous work the author found solutions to the Toda equations that
were expressed in terms of determinants of integral operators. Here it is 
observed that a simple variant yields solutions to the matrix Toda equations. 
As an application another derivation is given of a differential equation of Sato, Miwa 
and Jimbo for a particular Fredholm determinant.
\end{abstract}

During the last twenty years, beginning with \cite{MTW}, many connections have been 
established between determinants of integral operators and solutions of differential 
equations. The cited work concerned the integral operator $K$ on $L^2(\bR^+)$ with kernel
\[{e^{-{t\ov4}(u+u\inv+v+v\inv)}\ov u+v}.\]
It was shown that $\tau:=\log\,\det\,(I-\la^2K^2)$ has the representation
\bq\tau=-\hf\int_t^{\iy}s\,\Big(({d\ph\ov ds})^2-\sinh^2\ph\Big)\,ds,\label{taurep}\eq
where $\ph=\ph(t;\la)$ satisfies the differential equation
\bq{d^2\ph\ov dt^2}+{1\ov t}{d\ph\ov dt}=\hf\,\sinh 2\ph\label{phieq}\eq
with boundary condition
\[\ph(t;\la)\sim 2\la K_0(t)\ \ {\rm as}\ t\ra\iy.\]
(Here $K_0$ is the usual modified bessel function.) The differential equation for $\ph$,
the cylindrical sinh-Gordon equation, is reducible to a special case of the Painlev\'e III 
equation. This result was the first of
several in which special integral operators were shown to have determinants expressible
in terms of Painlev\'e functions. 

The proof in \cite{MTW} was combinatorial in nature
and quite difficult. Simpler proofs of a somewhat stronger result have been obtained since 
then. Note that differentiating (\ref{taurep}) twice and 
using the equation (\ref{phieq}) gives the equivalent relation
\bq{d^2\tau\ov dt^2}+{1\ov t}{d\tau\ov dt}=-\sinh^2\ph.\label{taueq1}\eq
It follows from results in \cite{B&LC} (see also \cite{TW}) that if we define
$\tau^{\pm}:=\log\,\det\,(I\pm\la K)$ then
\[{d^2\tau^{\pm}\ov dt^2}+{1\ov t}{d\tau^{\pm}\ov dt}={1-e^{\pm 2\ph}\ov4},\]
where $\ph$ solves (\ref{phieq}). Adding the two equations give (\ref{taueq1}). 

Subtracting the two equations and comparing with (\ref{phieq}) shows that 
\[\ph=\log\,\det\,(I+\la K)-\log\,\det\,(I-\la K)\]
solves (\ref{phieq}). Another proof of this fact 
was given in \cite{W}. Here 
families of operators $G_k$ (with $k\in\bZ$) depending on parameters $x$ and $y$ were 
produced such that the functions $q_k:=\log\,\det\,(I-G_{k+1})-\log\,\det\,(I-G_k)$ 
satisfy the Toda equations
\[{\pl^2 q_k\ov\pl x\pl y}=e^{q_k-q_{k-1}}-e^{q_{k+1}-q_k},\qquad k\in{\bZ}.\]
In a special case $\det\,(I-G_k)$ was a function of the product $xy$ and $G_k(t/4,t/4)$
was equal to $(-1)^k\,\la\,K$ with $K$ as given above. Equation (\ref{phieq}) 
followed from these
facts and the observation that $q_0=\ph,\ q_{-1}=q_1=-\ph$. Notice that these
solutions of the Toda equations are 2-periodic in the sense that $q_{k+2}=q_k$.

The purpose of this 
note is to give a ``Toda'' proof of a generalization of the first-cited result
which was established in \cite{SMJ}. Here a parameter $\th$ was introduced into the
kernel of $K$, so that it equals
\[\Big({u\ov v}\Big)^{\th/2}\,{e^{-{t\ov4}(u+u\inv+v+v\inv)}\ov u+v}.\]
It was shown that if we define 
\[\tau:=\log\,\det\,(I-\la^2K\,K')\]
($'=$transpose) then (\ref{taueq1}) holds, where $\ph$ now satisfies
\bq{d^2\ph\ov dt^2}+{1\ov t}{d\ph\ov dt}=\hf\,\sinh2\ph+
{\th^2\ov t^2}\,\tanh\ph\;{\rm sech}^2\ph\label{phieq2}\eq
with boundary condition
\[\ph(t;\la)\sim 2\la K_{\th}(t)\ \ {\rm as}\ t\ra\iy.\]
This can also be reduced to a special case of the Painlev\'e III equation.

Since the determinant of $I-\la^2K\,K'$ is equal to the determinant of the operator
matrix \mbox{\scriptsize$\ba I&\la K\\\la K'&I\ea$} it is not surprising that this fact can be proved by
extending the results of \cite{W} to obtain solutions of the 2-periodic {\it matrix} Toda
equations by means of operators with matrix-valued kernels.
Notice that in the scalar case described above if we set $Q_k:=e^{q_k}$ then the
Toda equations become
\bq{\pl\ov\pl y}\Big({\pl Q_k\ov\pl x}/Q_k\Big)={Q_k\ov Q_{k-1}}-{Q_{k+1}\ov Q_k}.
\label{QToda}\eq
The matrix Toda equations are the generalization of this given by
\bq{\pl\ov\pl y}\Big({\pl Q_k\ov\pl x}Q_k\inv\Big)=Q_k\,Q_{k-1}\inv
-Q_{k+1}\,Q_k\inv,\label{MToda}\eq
where the $Q_k$ are now matrix functions of $x$ and $y$. 

We shall now be more explicit about the relevant result of \cite{W} and its matrix
extension. Define $E(u):=e^{-(xu+yu\inv)}$
and let $p(u)$ be a suitable function on $\bR^+$. (It is only required that the
operators which occur are trace class.) Define $G$ to be the integral operator on 
$L^2(\bR^+)$ with kernel
\bq G(u,v)={p(u)\,E(u)\,p(v)\,E(v)\ov u+v},\label{Gk}\eq
set $G_k:=(-1)^kG$ and assume that the operators $I-G_k$ are invertible. Then a 
(clearly 2-periodic) solution of the Toda system (\ref{QToda}) is given by 
\bq Q_k={\det\,(I-G_{k+1})\ov\det\,(I-G_k)}.\label{detsol}\eq
Moreover we also have
\[Q_k=1+(-1)^k\Big(pE_0,\,(I-G_k)^{-1}\,pE_{-1}\Big),\]
where we define $E_i(u):=u^iE(u)$. 

An examination of the derivation of this reveals that, with only trivial changes, one can
establish the following matrix version: In the formula (\ref{Gk}) replace $p(u)$ and $p(v)$ 
by matrix functions $p(u)$ and $q(v)$, respectively. Then a solution to (\ref{MToda}) is
given by
\bq Q_k=I+(-1)^k\Big(qE_0,\,(I-G_k)^{-1}\,pE_{-1}\Big),\label{invsol}\eq 
where the inner product is interpreted as matrix multiplication
(in the order indicated) followed by integration. We also have
\bq \det\,Q_k={\det\,(I-G_{k+1})\ov\det\,(I-G_k)},\label{Mdet}\eq
which is the replacement of (\ref{detsol}).

Next we state a fact about these solutions which could easily have been derived in
\cite{W} but was not. This is that for the (scalar) solutions of
(\ref{QToda}) we have
\[-\pxy \log\,\det\,(I-G_k)={Q_k\ov Q_{k-1}}-1,\]
and more generally for the (matrix) solutions of (\ref{MToda}) we have
\bq-\pxy \log\,\det\,(I-G_k)=\tr(Q_k\,Q_{k-1}\inv-I).\label{pldetrep}\eq
At the end of this note we shall explain how this is proved.

We consider the special case where
\[p(u)=\ba f(u)&0\\0&g(u)\ea,\qquad q(u)=\ba 0&g(u)\\f(u)&0\ea.\]
For the present $f$ and $g$ are general although eventually they will be the functions 
$u^{\pm\th/2}$. We shall take $k=0$ and write $Q$ for $Q_0$. The 
kernel of $G$ is
\[G(u,v)=\ba 0&{f(u)\,E(u)\,g(v)\,E(v)\ov u+v}\\{g(u)\,E(u)\,f(v)\,E(v)\ov u+v}&0\ea
=\ba 0&A\\B&0\ea,\]
say. Since
\bq I\pm G=\ba I&\pm A\\\pm B&I\ea,\label{Gpm}\eq
we have
\bq\det\,(I\pm G)=\det\,(I-AB),\label{AB}\eq
so (\ref{Mdet}) gives 
\bq\det\,Q_k=1.\label{Qdet}\eq

From (\ref{Gpm}), the form of the matrices $p$ and $q$ and (\ref{invsol}) we easily
see that the diagonal elements of $Q_1$ are equal to those of $Q=Q_0$ 
while the off-diagonal elements are the negatives of each other. Similarly, interchanging 
$f$ and $g$
has the same effect on $I-G$ as left and right-multiplying by the matrix
\mbox{\scriptsize$\ba 0&1\\1&0\ea$} and from this it follows that the two diagonal entries of $Q$, as well
as the two off-diaginal entries, are obtained from each other by interchanging the
roles of $f$ and $g$. Denoting the effect of this interchange by a tilde, we see that
we may write our matrices as
\[Q=\ba 1+b&a\\\tl a&1+\tl b\ea,\qquad Q_1=\ba 1+b&-a\\-\tl a&1+\tl b\ea.\]
Observe that (\ref{Qdet}), which gives the identity
\bq b+\tl b+b\,\tl b=a\,\tl a,\label{uv}\eq
also gives
\[Q\inv=\ba 1+\tl b&-a\\-\tl a&1+b\ea,\qquad 
Q_1\inv=\ba 1+\tl b& a\\\tl a&1+b\ea.\]
And from these and (\ref{pldetrep}) with $k=0$ we obtain

\bq-\pxy \log\,\det\,(I-G)=4a\tl a.\label{Gdiff}\eq\sp

Let us see what the matrix Toda equations (\ref{MToda}) give. When $k=0$ the
equation is
\[\Pxy{Q}\,Q\inv+\Px{Q}\,\Py{Q\inv}=Q\,Q_1\inv-Q_1\,Q\inv.\]
Comparing the entries of these matrices gives the four equations (we use subscript
notation now for partial derivatives)
\[\hspace{-12em}\begin{array}{rl}
({\rm i})&b_{xy}(1+\tl b)-a_{xy}\tl a+b_x\tl b_y-a_x\tl a_y=0,\\
({\rm ii})&\tl b_{xy}(1+b)-\tl a_{xy}a+\tl b_xb_y-\tl a_xa_y=0,\\
({\rm iii})&a_{xy}(1+b)-ab_{xy}+a_xb_y-b_xa_y=4a(1+b),\\
({\rm iv})&\tl a_{xy}(1+\tl b)-\tl a\tl b_{xy}+\tl a_x\tl b_y-\tl b_x\tl a_y=4\tl a(1+\tl b).\\
\end{array}\]

Equations (i) and (ii) may be written
\[(b_x(1+\tl b)-a_x\tl a))_y=0,\ \ \ (\tl b_x(1+b)-\tl a_x a))_y=0\]
and since all our functions vanish as $y\ra +\iy$ we deduce
\bq b_x(1+\tl b)=a_x\tl a,\ \ \ \tl b_x(1+b)=\tl a_x a.\label{xders}\eq

We derive analogous identities for $y$-derivatives as follows. Denote by $T$ the 
unitary operator defined by $Th(u)=u\inv h(u\inv)$, and denote by a carat the effect
of the replacements $f(u)\ra f(u\inv),\ g(u)\ra g(u\inv)$. Then (we now display the
dependence of everything on the parameters $x$ and $y$) 
we find that $TG(x,y)T=\hat G(y,x),\ T(qE_0(x,y))=\hat q\hat E_{-1}(y,x),\ 
T(pE_{-1}(x,y))=\hat p\hat E_0(y,x)$. Thus, if we set
\[U:=\Big(qE_0,\,(I-G)^{-1}\,pE_{-1}\Big),\ \ \
V:=\Big(qE_{-1},\,(I-G)^{-1}\,pE_0\Big),\]
then $U(x,y)=\hat V(y,x)$. On the other hand, the symmetry of $G$ (the fact that its kernel
satisfies $G(u,v)'=G(v,u)$) implies that
$V'=(p'E_0,\,(I-G)\inv\,q'E_{-1}).$
We have, using the same tilde notation as before and setting $S:=$\mbox{\scriptsize
$\ba 0&1\\1&0\ea$},
\[p'=\tl q S,\qquad q'=S\tl p,\qquad SGS=\tl G,\]
and from this we deduce that $V'=\tl U$. Combining this with the already established
$U(x,y)=\hat V(y,x)$ we deduce $\tl U(x,y)=\hat U'(y,x)$, in other words
\[a(x,y)=\hat a(y,x),\ \ \tl a(x,y)=\tl{\hat a}(y,x),\ \  b(x,y)=\tl{\hat b}(y,x),\ \ 
\tl b(x,y)=\hat b(y,x).\]
Combining these with (\ref{xders}) for the operator $\hat G$ we obtain
\[\tl b_y(1+b)=a_y\tl a,\ \ \ b_y(1+\tl b)=\tl a_y a.\]

Eliminating $b_{xy}$ and $\tl b_{xy}$ from equations (i) and (iii), and (ii) and (iv),
respectively and using our formulas for the 
derivatives of $b$ and $\tl b$ as well as (\ref{uv}) we find the equations
\bq 
a_{xy}={\tl a\ov 1+a\tl a}\,a_x\,a_y+4\,a\,(1+a\tl a),\ \ \ \ 
\tl a_{xy}={a\ov 1+a\tl a}\,\tl a_x\,\tl a_y+4\,\tl a\,(1+a\tl a).
\label{ueq}\eq\sp

These equations hold whatever the functions $f$ and $g$. We now use them to obtain the cited
result of \cite{SMJ}. By (\ref{AB}) we see that the determinant in question is equal
to $\det\,(I-G)$ evaluated at $x=y=t/4$ in the case where
\[f(u)=\sqrt{\la}\,u^{\th/2},\ \ g(u)=\sqrt{\la}\,u^{-\th/2}.\]
Observe first that $\hat a=\tl a$ in this case, so that $\tl a(x,y)=a(y,x)$. We now 
show that
\bq a(x,y)=\Big({x\ov y}\Big)^{\th/2}\,a(\sqrt{xy},\sqrt{xy}),\ \ \ 
\tl a(x,y)=\Big({y\ov x}\Big)^{\th/2}\,\tl a(\sqrt{xy},\sqrt{xy}).\label{arep}\eq
For this we take any $r>0$ and use the unitary operator $T$ now
defined by $Th(u)=r^{\hf}h(ru)$. Denote now by a carat the result of the replacement 
$(x,\,y)\ra(rx,\,y/r)$. Since $TGT=\hat G$ and
\[T(qE_0)=r^{\hf}\ba r^{-\th/2}&0\\0&r^{\th/2}\ea\,q\hat E_0,\ \ \ 
T(pE_{-1})=r^{-\hf}p\hat E_{-1}\ba r^{\th/2}&0\\0&r^{-\th/2}\ea,\]
we deduce
\[Q=\ba r^{-\th/2}&0\\0&r^{\th/2}\ea\,\hat Q\,\ba r^{\th/2}&0\\0&r^{-\th/2}\ea,\]
which gives the asserted identities upon setting $r=\sqrt{y/x}$.

We also deduce from $TGT=\hat G$ in the same way that $\det\,(I-G)$ is a function of 
$xy$, and we shall eventually set $x=y=t/4$. Since for a function of $t=4\sqrt{xy}$
\[\pxy=4\,({d^2\ov dt^2}+t\inv{d\ov dt}),\]
the left side of (\ref{taueq1}) equals $1/4$ times the left side of 
(\ref{Gdiff}) evaluated at $x=y=t/4$.
Thus if we set $c(t):=a(t/4,t/4)=\tl a(t/4,t/4)$ and define $\ph$ by $\sinh\ph=c$,
then (\ref{taueq1}) holds and it remains to verify (\ref{phieq2}). Using (\ref{arep}) 
we find that either equation in (\ref{ueq}) becomes at $x=y=t/4$
\[{d^2 c\ov dt^2}+{1\ov t}{dc\ov dt}={c\ov 1+c^2}\Big({dc\ov dt}\Big)^2+c(1+c^2)+
{\th^2\ov t^2}\Big(c-{c^3\ov 1+c^2}\Big),\]
and (\ref{phieq2}) follows upon substituting $c=\sinh\ph$.\sp

\noi{\bf Remark}. In \cite{B&LC} differential identities were found, by
different methods, for the quantities we called $a,\ \tl a,\ b,\ \tl b$.
These identities do not seem to give our equations (\ref{ueq}). A general result was also
stated there which would imply in particular that (\ref{phieq}) holds rather than 
(\ref{phieq2}) for the operator kernel with general $\th$. The authors are aware of the 
error in their paper and plan to publish an erratum.\sp

\begin{center}{\bf Appendix}\end{center}

We derive (\ref{pldetrep}) here. Taking the logarithmic derivative of (\ref{Mdet}) with 
respect to $x$ gives
\[\tr \Big({\pl Q_k\ov\pl x}Q_k\inv\Big)={\pl\ov\pl x}\log\,\det\,(I-G_{k+1})-
{\pl\ov\pl x}\log\,\det\,(I-G_k),\]
and so taking traces in (\ref{MToda}) gives
\[\pxy\log\,\det\,(I-G_{k+1})-\pxy\log\,\det\,(I-G_k)=
\tr(Q_k\,Q_{k-1}\inv-Q_{k+1}\,Q_k\inv).\]
Suppose it were true (which it certainly is not) that $G_k\ra0$ in trace 
norm and $Q_k\ra I$  as $k\ra+\infty$. Then replacing $k$ successively by $k,\ k+1,\ \cdots$ 
in the above relation and adding would give (\ref{pldetrep}).

In order to make this argument work we use a family
of operator solutions to (\ref{QToda}) depending on a parameter $\om$, these also being 
special cases of those derived in \cite{W}. We assume that $\om$ belongs to
\[\Om:=\{\om\in\bC\backslash\bR^+:\ \Re\,\om<1,\ \Re\,\om\inv<1\},\] 
set $E(\om,u):=e^{-[(1-\om\inv)xu+(1-\om)yu\inv]/2}$,
define $G$ to be the operator on $L_2(\bR^+)$ with kernel
\[{p(u)\,E(\om,u)\,p(v)\,E(\om,v)\ov u-\om v},\]
and set $G_k:=\om^kG$. Then
\[Q_k=1+\om^k\Big(pE_0,\,(I-G_k)^{-1}\,pE_{-1}\Big)\]
(where we now define $E_i(u):=u^iE(\om,u)$) satisfies (\ref{QToda}) and (\ref{detsol})
whenever these make sense, i.e., when the operators $I-G_k$ that appear in the
expressions are invertible. In the matrix version the factors $p(u)$ and $p(v)$
are replaced by matrix functions $p(u)$ and $q(v)$, the constant 1 in the definition
of $Q_k$ is replaced by $I$, and (\ref{MToda}) and (\ref{Mdet})
hold. Notice that we are interested in the case $\om=-1$.

Let $W$ be any open set whose closure is a compact subset of
$\{\om\in\Om:\ |\om|<1\}$.
Then for some $k'$ all the operators $G_k$ with $k\geq k'$ will have norm less than
1 when $\om\in W$ and so the $I-G_k$ will be invertible. (We think of $x$ and $y$ as 
lying in fixed intervals bounded away from 0.) Now let $k_0$ be arbitrary. For fixed 
$x$ and $y$, removing a 
finite set from $W$ will ensure that all $I-G_k$ with $k\geq k_0$ are invertible.
If $x$ and $y$ are confined to sufficiently small intervals there will still be a
non-empty open subset $W_0$ of $W$ such that all $I-G_k$ with $k\geq k_0$ and $\om\in W_0$
are invertible. Moreover since $|\om|<1$ in $W_0$ it is clear that
$G_k\ra0$ in trace norm and $Q_k\ra I$ as $k\ra+\infty$,
so the argument given above shows that (\ref{pldetrep}) holds in this case for all
$k\geq k_0$. But both
sides of the identity are analytic functions of $\om\in\Om$ and taking a suitable path
in $\Om$ running from a point in $W_0$ to $\om=-1$ we deduce (\ref{pldetrep})
for $\om=-1$, in other words for our given operator.

\begin{center}{\bf Acknowledgement} \end{center}
This work was supported by National Science Foundation 
grant DMS-9424292.

\end{document}